 \documentclass[preprint,12pt]{elsarticle}
\usepackage{epsfig}

\usepackage{amssymb}
\begin{document}

\begin{frontmatter}



\title{Experimental results and constitutive modelling for tungsten and tantalum at high strain rates and very high
temperatures\tnoteref{t1}}
\tnotetext[t1]{Work supported by Science and Technology Facilities Council (UK).}

\author[shf]{G. P. \v {S}koro\corref{cor1}}
\ead{g.skoro@sheffield.ac.uk}

\author[ral]{J. R. J. Bennett}

\author[ral,hud]{T. R. Edgecock}

%\author[shf]{C. N. Booth}

\cortext[cor1]{Corresponding author}
\address[shf]{Department of Physics and Astronomy, University of Sheffield, Sheffield S3 7RH, UK}
\address[ral]{STFC, Rutherford Appleton Laboratory, Chilton, Didcot, Oxon OX11 0QX, UK}
\address[hud]{University of Huddersfield, Queensgate, Huddersfield HD1 3DH, UK}
%\address[war]{Department of Physics, University of Warwick, Coventry CV4 7AL, UK}

%%\author{}

%%\address{}

\begin{abstract}
Recently reported results of the high strain rates, high temperature
measurements of the yield stress of tungsten and tantalum have been analyzed.
The highest temperature reached in the experiment, based on heating and stressing
a thin wire by a fast, high current pulse, was $2250\ ^{0}$C and
$2450\ ^{0}$C, for tantalum and tungsten, respectively. The strain-rates in
both the tungsten and tantalum tests were in the range from $500$ to $1500\ s^{-1}$.
The parameters for the constitutive equation developed by Zerilli and
Armstrong have been determined from the experimental data and the results have been compared with the data obtained at
lower temperatures.

\end{abstract}


\end{frontmatter}

% \linenumbers

\section{Introduction}
\label{}
Tungsten, tantalum and their alloys are the prefered candidates for many
high strain rate applications, for example kinetic energy 
penetrators \cite{Pen, Extr}.
Tungsten has been used for many decades in high
temperature applications. Even so, there is a lack of experimental data on 
the thermomechanical response of refractory metals to high strain rate
effects at temperatures higher than $1000\ ^{0}$C.

The behaviour of materials under these conditions is of great interest from
phenomenological point of view, for example for testing the consistency of
the different constitutive models. In addition to this, the materials used
in the target systems in the next generation of high power particle accelerators
will be more and more frequently exposed to the combination of high stresses,
high strain rates and high temperatures.
In order to estimate the lifetime of the target and target system components
in this
environment, the candidate materials must be tested under these extreme conditions.

As part of the UK programme of high power target developments for a
Neutrino Factory \cite{Nufact} a new dynamic method for thermomechanical characterization
of the candidate materials (tungsten and tantalum in this case) has been developed
\cite{Jnm, Nim}. The method is based on heating and stressing
a thin wire (less than $1$ mm diameter) of the candidate material by a
fast ($\sim 1 \mu$s long), high current (up to $9200$ A) pulse. This paper is
a continuation of the study presented in \cite{Nim} where the yield strength of
tungsten and tantalum has been measured at strain rates from $500$ to $1500\ s^{-1}$
and at temperatures much higher than previously recorded in the literature.
The highest temperature reached in the experiment was $2250\ ^{0}$C and
$2450\ ^{0}$C, for tantalum and tungsten, respectively. It should be noted that
reference \cite{Nim} has a different aim and scope and that this paper is focused
on the analysis of the new yield strength data, constitutive modelling and
comparison with previous results for tungsten and tantalum at elevated temperatures.

%Because of this and for the sake of completeness, description of the
%experimental procedure have been briefly reproduced from \cite{Jnm, Nim} but with
%additional details that are more relevant for the purpose of this analysis.

The rest of the paper is organized as follows. The
description of the experimental procedure and results
are given in Section 2. 
Section 3 describes the constitutive modelling procedure with special attention paid
to the Zerilli and Armstrong model \cite{ZA87}.
The results are discussed in Section 4, 
followed by the summary in Section 5. 

\section{Experimental procedure and results}
A thin wire is necessary to allow the current to diffuse into the centre of the wire in a sufficiently short time 
to generate the required thermal stress. 
The tungsten and tantalum wires with diameters from $0.5$ mm to $0.8$ mm have
been made by standard powder metallurgy methods - pressing, sintering,
forging and finally drawing. The purity of the wires used in
the tests was at least $99.9 \%$. 
The tungsten wires have been additionally stress relieved at about $400\ ^{0}$C.

To determine the yield strength of the tantalum and tungsten wires the pulse amplitude of the current in the wire was 
increased in steps at a fixed temperature.  
The current was kept constant for 3-5 minutes before taking the next increase. This was continued until 
the wire was observed to start bending or kinking. The surface motion of the test wire during 
the pulsing is measured by a laser Doppler vibrometer (LDV) from Polytec \cite{Poly}.
The radial surface velocity measured by the LDV (see, for example, Figure 1),
is used to extract the strain rate during the test.
The optical sensor head of the LDV includes an integrated fast camera that is used to visually
monitor the strain of the wire. In addition, it was noticed that the LDV radial
velocity 
signal would become very noisy as one approached the first signs of plastic deformation, eventually resulting in the 
loss of any coherent vibration signal. This monitoring and the change of the
quality of the LDV signal were the main indicators that the wire is near or at the bending
point.

The bent wire was then replaced with a new, straight
sample and the experiment was repeated at a new temperature and current. In some cases, the
procedure was changed so the current was kept constant and the temperature was changed by
adjusting the pulse repetition rate. In order to estimate the systematic
uncertainties, at some temperatures there were several tests and it was found
that the results were repeatable to
within $\pm5\%$ of the current. An optical pyrometer was used to measure the temperature of the wire at the same point measured by the 
vibrometer.

The stress in the wire is not directly measured, but it is proportional to the
square of the current and can be modelled by using modern finite element codes.
The first step is to measure the current pulse in the wire, then to calculate 
the current density as a function of time and radius using the solution to the
diffusion equation \cite{De25}. 
Knowing the current density, the temperature rise is calculated and finally 
the finite element code LS-DYNA \cite{Dyna} is used to calculate the equivalent
(von Mises) stress as a function of time and radius (more details can be
found in \cite{Jnm,Nim,Gor}). As the stress in the
wire is a result of calculation it is important to benchmark the results obtained.
Figure 1
shows the comparison between measured radial surface velocity of the tungsten wire at different
temperatures and the
LS-DYNA results \cite{Nim}. The agreement between experiment and simulation
is very good and this can be taken as a proof that the corresponding calculated
stresses are correct. 

Figure 2 shows the stress at which the tantalum and tungsten wires reached
the yield point as a function of temperature. The lower
edge of the wide bands indicates the stress at which the wire still appeared undeformed
and the upper edge 
the stress at which the wire started to bend. For tantalum, bands are shown
for wires of $0.5$ and $0.8$ mm diameter. In both cases, the current was increased
in the same, fixed steps. As the stress is inversely proportional to the square
of the diameter, the band for the $0.5$ mm diameter tantalum wire in Figure 2 is much wider
than for the $0.8$ mm diameter wire. In the case of tungsten, $0.5$ mm diameter wires have
been tested. The highest temperature reached in the experiment was $2250\ ^{0}$C and
$2450\ ^{0}$C, for tantalum and tungsten, respectively. On the low side, it was not
possible to go below $1150\ ^{0}$C (tantalum) and
$1450\ ^{0}$C (tungsten) as the minimum temperature is limited by maximum
available current. 

The peak strain rate is determined by dividing the measured
peak radial velocity (see Figure 1) by the wire radius. The peak strain rates for
the measurements of both tantalum and tungsten cover the range $(500-1500)\ s^{-1}$. 
The characteristic values of the strain rate are indicated at various points in 
Figure 2. In this
experiment the strain rates decrease at high temperatures. 
The strain of the wire during the pulsing was monitored by a fast camera. 
These observations were compared to the change in the
quality of the LDV signal and it was found that the strain at the yield point was
usually between $2\%$ and $4\%$. Those values have been used also in the
constitutive modelling (see Section 3).

Once a wire has begun to distort the stress in the wire increases due to the curvature of
the wire. Also accompanying the bending, there is usually stretching and thinning of the
wire at the hottest point,
so the stress and temperature of the wire increases locally. This all happens within a
few pulses and 
is soon followed by severe bending and/or breaking, so that it is difficult to measure
the temperature, diameter and to calculate 
the true stress in the wire as the plastic strain increases. 
So, in this experiment it was not possible to measure the 'complete' stress-strain curve
but only the yield point.
    
\section{Constitutive modelling}

Zerilli and Armstrong \cite{ZA87} proposed the following constitutive (Z$\&$A)
equation for the
flow stress in body centred cubic (bcc) metals as a function of temperature $T$,
strain $\epsilon$ and strain rate $\dot{\epsilon}$:
\begin{equation}
\sigma  = C_1e^{(-C_2T+C_3Tln\dot{\epsilon})}+C_4+C_5\epsilon^n+kd^{-1/2}.
\end{equation}
$C_1$ to $C_5$, $n$ and $k$ are material parameters, while $d$ is the grain size.
If the precise information about grain size is not available (as in our case), the
usual procedure \cite{Ch95,Extr,ZA90,Chgr} is to incorporate
the grain size term into the $C_4$ parameter because both
are independent of
temperature, strain and strain-rate:
\begin{equation}
\sigma  = C_1e^{(-C_2T+C_3Tln\dot{\epsilon})}+C_4+C_5\epsilon^n.
\end{equation}
The first term in this equation combines effects of 'thermal softening'
(reduction of yield strength with increasing temperature) and
'strain-rate hardening' (rise of yield strength with increasing strain-rate).
This term does not depend on strain. The parameters $C_5$ and $n$ describe the strain hardening changes.
In our case, $C_4$ and $C_5$ terms are practically constants (see the
previous section) but they are
left to vary during the fitting procedure.
 
The CERN computer package for function minimization MINUIT \cite{Min} was used to vary the
parameters $C_1$ to $C_5$ and $n$ and to minimise the
sum of the squares of the deviations ($\chi^2$) between the values obtained from
Z$\&$A equation and experimental (fitting) points.
The stress value, for each measured set of
($T$, $\epsilon$, $\dot{\epsilon}$), was chosen to lie in the middle
between the corresponding upper and lower stress  values (see Figure 2). The weighted least square
fit was used, where the weight of the
fitting point, $i$, is $w_i=1/{\sigma_i}^2$, $\sigma_i$ being the half of the difference between upper and lower
edge stress values at particular temperatures.  

The best fit of the tantalum data is shown in Figure 3. One can note that the $0.5$
and $0.8$ mm data are combined into one set for fitting purposes. Due to the weighted fitting
method the fitting curve has a characteristic shape in the region where the
results overlap each others. Figure 4 shows the best fit for the tungsten data.
In both cases the normalized $\chi^2$ value is close to 1.

The corresponding optimized parameters values for tungsten and tantalum are
presented in Table 1. The values of the $C_1$, $C_2$ and $C_3$ parameters
for tungsten and tantalum are very similar. This lead to the conclusion that
the yield strength dependance on temperature and strain-rate (the slope of the curve)
for tantalum and tungsten at very high temperatures and strain rates at
around $1000\ s^{-1}$ is similar.

Finally, using an identical fitting procedure the experimental data was fitted with
Johnson-Cook model \cite{JC}, but this model
was found to be inadequate for describing our experimental data. This is in
agreement with previous conclusions that the Z$\&$A
model is much better equipped to capture strain-rate dependence and thermal
softening behaviour of bcc metals \cite{Extr, Chgr}.

\section{Discussion}
\subsection{Tungsten}
Table 2 shows the values of the parameters of the Z$\&$A equation obtained in this work and
previous experiments with tungsten samples. 

A limited data set exists for pure tungsten behaviour at elevated temperatures and
high strain-rates. Chen and Grey \cite{Ch95}
have used commercially pure rolled tungsten plate for high strain-rate ($0.001-4000\ s^{-1}$)
tests at temperatures up to $1000\ ^{0}$C. Their fitting procedure resulted in a set of Z$\&$A parameters
denoted as 'CG' in Table 2 (and Figure 5 - see below).

Lennon and Ramesh \cite{Extr} have measured and fitted ('LR' set in Table 2)
the stress-strain curve of the samples made from heavily deformed extruded tungsten rod.
The commercially obtained extruded rod had
undergone about $80\%$ reduction in area (about $60\%$ effective
strain). The samples were tested in the strain-rate range of $(0.001 - 7000)\ s^{-1}$ and at
temperatures ranging from $27$ to $800\ ^{0}$C.

Tungsten samples for the tests performed by Dummer et al. \cite{Dumm} were subjected to different
heat treatments. The as-received tungsten were made by pressing the tungsten powder
at $1200\ ^{0}$C and then sintered at $2600\ ^{0}$C. The portions of the final samples
were annealed at $1750\ ^{0}$C, $2600\ ^{0}$C and $2800\ ^{0}$C. The experiments were
performed at room temperature only with the strain-rate ranging from $0.001$ to $4000\ s^{-1}$.
The parameters of the Z$\&$A equation obtained for as-received tungsten samples were practically
the same as in the Chen and Grey experiment. In the case of annealed tungsten the yield stress was
reduced by about $40\%$. 

The common factor of these sets of tests is that the thermomechanical properties of
tungsten were measured at temperatures well below the temperature region explored
in this work. In an attempt to compare our new results with these, the tungsten yield stress
was calculated using the Z$\&$A constitutive
equation for the strain rate of $1000\ s^{-1}$ and strain of $3\%$
(characteristic values in this experiment) and plotted in Figure 5 as a function
of temperature. All parameter sets from Table
2 have been used and the results based on parameters from \cite{Ch95,Extr} are extrapolated
to cover the temperature range of our experiment. Visual inspection of Figure 5 reveals an
excellent agreement between this work and the 'CG' parametrization of the 
Z$\&$A equation while the 'LR' parametrization has a different slope.

\subsection{Tantalum}
Table 3 shows the values of the parameters of the Z$\&$A equation obtained in this work and
previous experiments with tantalum samples.

One of the first applications of the Z$\&$A model was in the analysis of tantalum experimental results
when Zerilli and Armstrong \cite{ZA90} fitted Hoge and Mukherjee experimental
data \cite{HM}. Hoge and Mukherjee have tested $99\%$ pure, fully recrystallized, tantalum
in the strain-rate range of $(0.00001 - 20000)\ s^{-1}$ and at
temperatures ranging from $-249$ to $527\ ^{0}$C. The corresponding set of obtained parameters
is denoted as 'ZA-HM' in Table 3 (and Figure 6 - see below). 

Chen and Grey have fitted the same 
experimental results \cite{HM} by optimizing the fit for the entire range
of data (parameter set 'CG-HM' in Table 3). They have also performed their own experiment described 
in the same paper \cite{Ch95} where commercially pure tantalum was tested in the strain-rate range
of $(1500 - 5000)\ s^{-1}$ and at temperatures up to $1000\ ^{0}$C. One of the conclusions of the
experiment was that the new data are very different from the Hoge and Mukherjee data\footnote{In Hoge and
Mukherjee experiment a different experimental technique (tension) was used. In all other experiments
cited in this paper a Split-Hopkinson pressure bar technique has been used \cite{SH}.}.
The set of Z$\&$A parameters that corresponds to Chen and Grey experimetal results \cite{Ch95} are shown
in Table 3 (denoted as 'CG').

In another Chen and Grey experiment \cite{Chgr} commercially pure, triple electron beam melted, vacuum
annealed tantalum plates were tested in the strain-rate range of $(0.001 - 4000)\ s^{-1}$ and at
temperatures ranging from $-196$ to $1000\ ^{0}$C.
Tantalum samples in those tests were prepared by melting large ingots, then
forging them into billets which were then annealed and cut prior to cross rolling. Finally, the
plates were straight rolled in the final finishing passes. The fitting of corresponding experimental
data resulted in a set of Z$\&$A parameters denoted as 'CG-AP' in Table 3.

As in the tungsten case, the common thing for all those experiments is that the thermomechanical
properties of tantalum were measured at temperatures well below those explored
in this work. In order to compare our experimental results with previous ones,
the tantalum yield stress
was calculated using the Z$\&$A constitutive
equation with a strain rate of $1000\ s^{-1}$ and strain of $3\%$. All parameter sets from Table
3 have been used and the results based on parameters from \cite{Ch95,ZA90,Chgr} are extrapolated
to cover temperature range in our experiment (see Figure 6). 

Unlike the tungsten case, there is a clear difference in Figure 6 between each of the extrapolated
parametrizations and the new experimental results. On the other hand, the extrapolated results differ
significantly between themselves (up to a factor of $2$) in the temperature region explored in our
experiment, but there is a common feature for all of them: a very weak 
temperature dependence of the yield
strength. However, our experimental results show a much stronger temperature sensitivity, and it is
interesting that they almost fully cover the range between the two extreme extrapolated curves shown
in Figure 6.

\section{Summary}
An analysis of the extensive set of high strain rates measurements of the
yield strength of tungsten and
tantalum at the record high temperatures (in the range of $1350-2700$ K) has been
performed. The strain-rates in
both the tungsten and tantalum tests were in the range from $500$ to $1500\ s^{-1}$.
The parameters for the constitutive equation developed by Zerilli and
Armstrong have been determined from the experimental data. Also, it has been found that the
Johnson-Cook model is inadequate for describing our test results.
The obtained parametrization of the Zerilli-Armstrong model has been compared with the
extrapolated parametrizations obtained in the tests at lower temperatures. 
It has been found that in the tungsten case
our results and the extrapolated Chen-Gray parametrization agree very well.
However, our experimental results for tantalum show a much stronger temperature sensitivity 
than the extrapolated
parametrizations obtained from the tantalum tests at lower temperatures.

%\section{Acknowledgements} 

%    The authors would like to thank members of the ISIS management and the Pulsed Power
%    Section of the Electrical Engineering Group of ISIS for their support
%    in giving access to the pulsed power supply and general
%    help with the experiment.

% \appendix

% \section{Acknowledgements}

%% References


\bibliographystyle{model1a-num-names}
\bibliography{<your-bib-database>}



%------------------------------------------
\newpage
\vspace*{5.5cm}

\begin{center}

\vskip 0.cm
{Table 1. Z$\&$A model parameters for tungsten and tantalum.}
\end{center}
\vskip -0.5cm

\begin{table}[htb]
   \centering
%   \caption{Results of the measurements of the surface roughness of our test samples.}
   \begin{tabular}{|c|c|c|}
%       \toprule
\hline\hline
{Parameter} &  {Tungsten} & {Tantalum} \\         
& {(W)} & {(Ta)}\\ \hline\hline
%       \midrule
C$_1$ [MPa] &  4711$\pm$261  &  4166$\pm$354      \\ \hline
C$_2$ [K$^{-1}$] &  (5.97$\pm$0.28)x10$^{-3}$ &  (7.89$\pm$0.41)x10$^{-3}$\\ \hline
C$_3$ [K$^{-1}$] &  (6.24$\pm$0.42)x10$^{-4}$ &  (8.04$\pm$0.38)x10$^{-4}$\\ \hline
C$_4$ [MPa] & 94$\pm$8 & 7.5$\pm$0.9        \\ \hline
C$_5$ [MPa] & 133$\pm$20   & 381$\pm$40     \\ \hline
n &  0.51$\pm$0.07  & 0.56$\pm$0.08      \\ \hline\hline
%       \bottomrule
   \end{tabular}
\end{table}

%------------------------------------------
%------------------------------------------
\newpage
\vspace*{5.5cm}

\begin{center}

\vskip 0.cm
{Table 2. Comparison between this work and previous results on Z$\&$A model parameters
for tungsten.}
\end{center}
\vskip -0.5cm

\begin{table}[htb]
   \centering
%   \caption{Results of the measurements of the surface roughness of our test samples.}
   \begin{tabular}{|c|c|c|c|}
%       \toprule
\hline\hline
&  {CG} & {LR} & {This} \\         
{Parameter} & {\cite{Ch95}} &  {\cite{Extr}} &  {work}\\ \hline\hline
%       \midrule
C$_1$ [MPa] & 3000 & 2749 &  4711        \\ \hline
C$_2$ [K$^{-1}$] &  2x10$^{-3}$ & 2.25x10$^{-3}$ & 5.97x10$^{-3}$\\ \hline
C$_3$ [K$^{-1}$] &  1x10$^{-4}$ & 9.05x10$^{-5}$ &  6.24x10$^{-4}$\\ \hline
C$_4$ [MPa] & 0 & 49.91 & 94.37        \\ \hline
C$_5$ [MPa] & 800 & 194.5 & 133        \\ \hline
n & 0.6 & 0.0505 & 0.512        \\ \hline\hline
%       \bottomrule
   \end{tabular}
\end{table}

%------------------------------------------
%------------------------------------------
\newpage
\vspace*{5.5cm}

\begin{center}

\vskip 0.cm
{Table 3. Comparison between this work and previous results on Z$\&$A model parameters
for tantalum.}
\end{center}
\vskip -0.5cm

\begin{table}[htb]
   \centering
%   \caption{Results of the measurements of the surface roughness of our test samples.}
   \begin{tabular}{|c|c|c|c|c|c|}
%       \toprule
\hline\hline
& {ZA-HM} & {CG-HM} & {CG} &  {CG-AP} & {This}\\         
{Parameter} & {\cite{ZA90}} & {\cite{Ch95}}& {\cite{Ch95}} & {\cite{Chgr}} & {work}\\ \hline\hline
%       \midrule
C$_1$ [MPa] & 1125 & 1200 & 975 & 1750& 4166        \\ \hline
C$_2$ [K$^{-1}$] & 5.35x10$^{-3}$ &  6.0x10$^{-3}$ & 4.5x10$^{-3}$ & 9.75x10$^{-3}$ & 7.89x10$^{-3}$\\ \hline
C$_3$ [K$^{-1}$] & 3.27x10$^{-4}$ & 3.875x10$^{-4}$ & 2.75x10$^{-4}$ & 6.75x10$^{-4}$ & 8.04x10$^{-4}$\\ \hline
C$_4$ [MPa] & 30 & 25 & 40& 140 & 7.5        \\ \hline
C$_5$ [MPa] & 310 & 310& 525& 650 & 381        \\ \hline
n & 0.44 & 0.44 & 0.5& 0.65 &  0.56        \\ \hline\hline
%       \bottomrule
   \end{tabular}
\end{table} 

%----------------------------------------
\newpage

%\vskip 6.5cm
\vspace*{8.5cm}

\begin{center}
\hspace*{-9.cm}
\parbox{2cm}{\epsfxsize=2.cm \epsfysize=2.cm \epsfbox[00 00 90 90]
{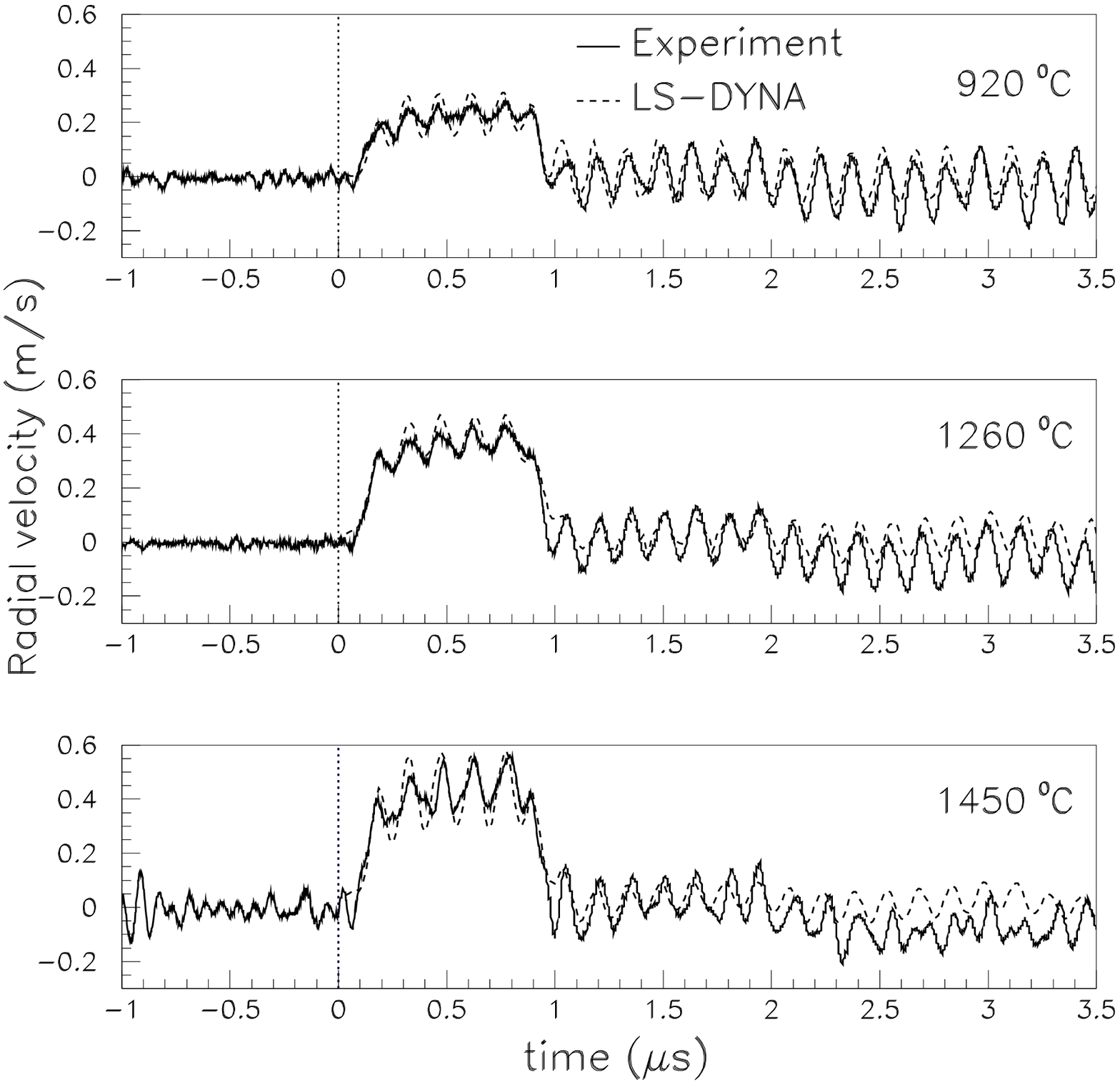}{}}

\vskip 0.3cm
{Figure 1. The measured and calculated radial velocity of a 0.5 mm diameter tungsten
wire at peak temperatures of $920$, $1260$ and $1450^0$C \cite{Nim}.}
\end{center}
\vskip 0.5cm

%------------------------------------------
\newpage

\vspace*{8.5cm}

\begin{center}
\hspace*{-9.cm}
\parbox{2cm}{\epsfxsize=2.cm \epsfysize=2.cm \epsfbox[00 00 90 90]
{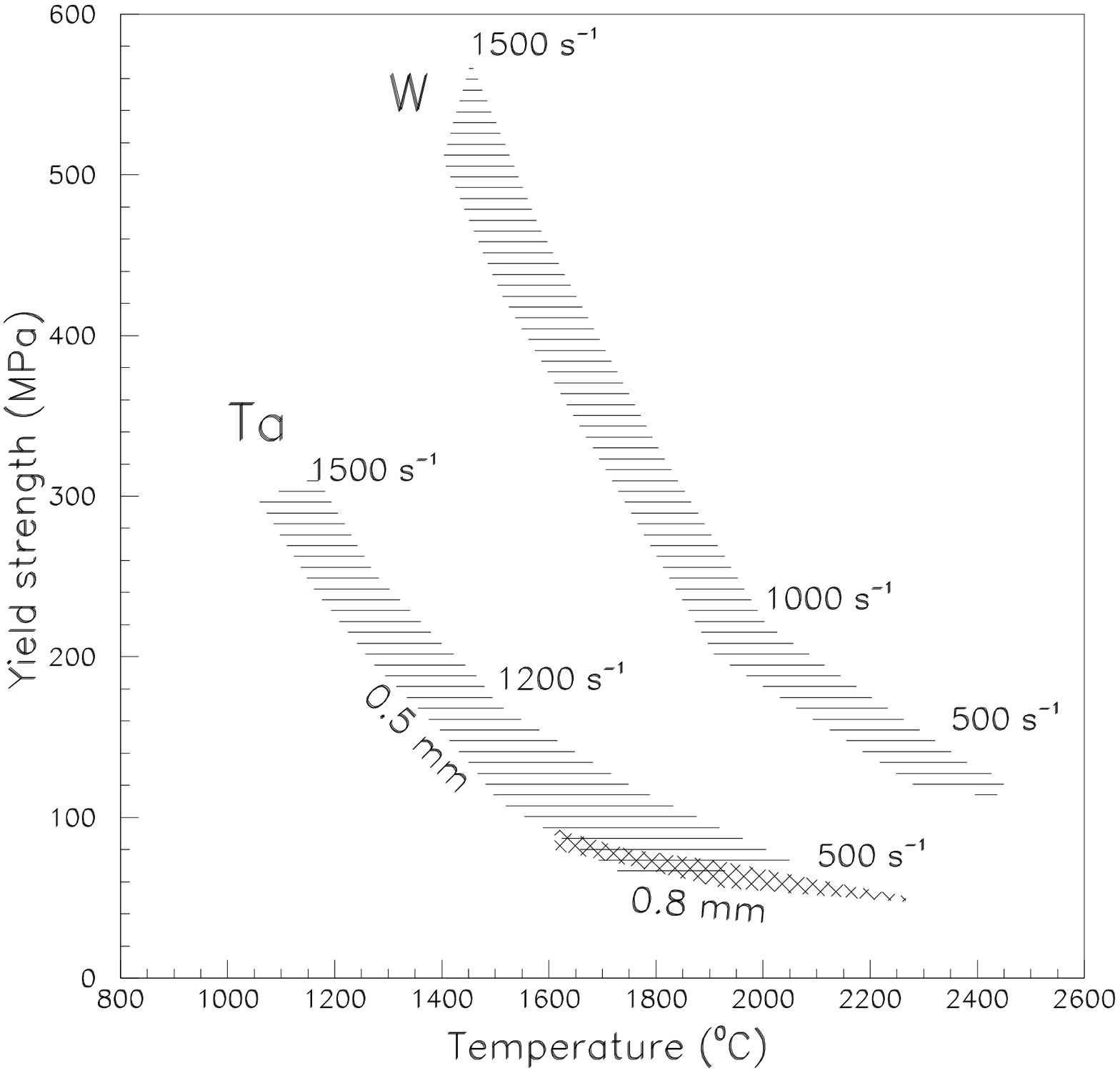}{}}

\vskip 0.3cm
{Figure 2. The yield strength versus peak temperature for tantalum wires of $0.5$ and
$0.8$ mm diameter and for tungsten wires of $0.5$ mm diameter \cite{Nim}. The upper edge of the
bands indicates the stress at which the wire started to bend and the lower edge indicates where
the wire was not deformed.
The characteristic strain rate values are indicated.}
\end{center}
\vskip 0.5cm
 
%------------------------------------------
\newpage
\vspace*{8.5cm}

\begin{center}
\hspace*{-9.cm}
\parbox{2cm}{\epsfxsize=2.cm \epsfysize=2.cm \epsfbox[00 00 90 90]
{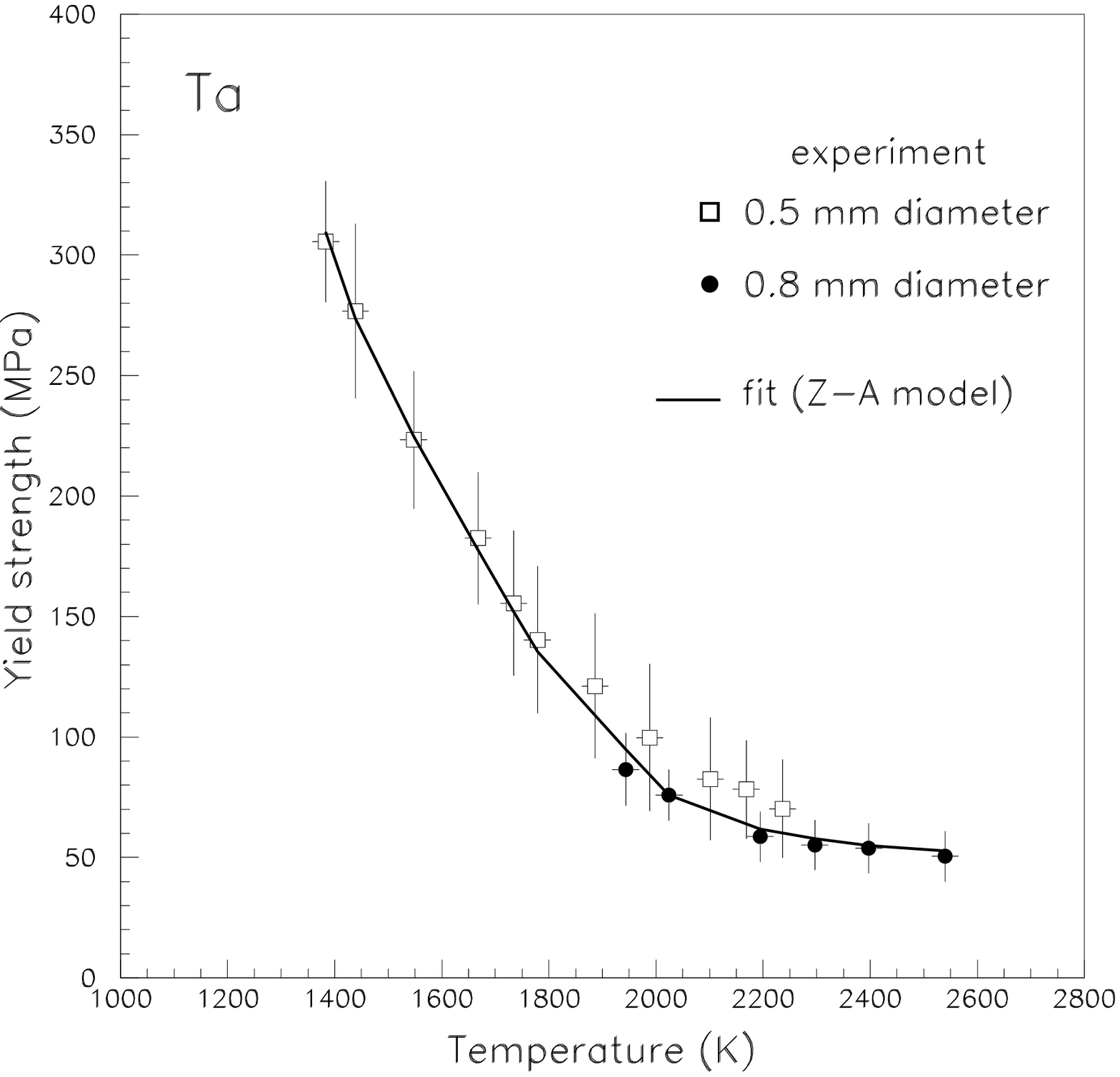}{}}

\vskip 0.3cm
{Figure 3. Experimental data on tantalum yield strength and the best fit
based on Z$\&$A model.}
\end{center}
\vskip 0.5cm

%------------------------------------------
\newpage
\vspace*{8.5cm}

\begin{center}
\hspace*{-9.cm}
\parbox{2cm}{\epsfxsize=2.cm \epsfysize=2.cm \epsfbox[00 00 90 90]
{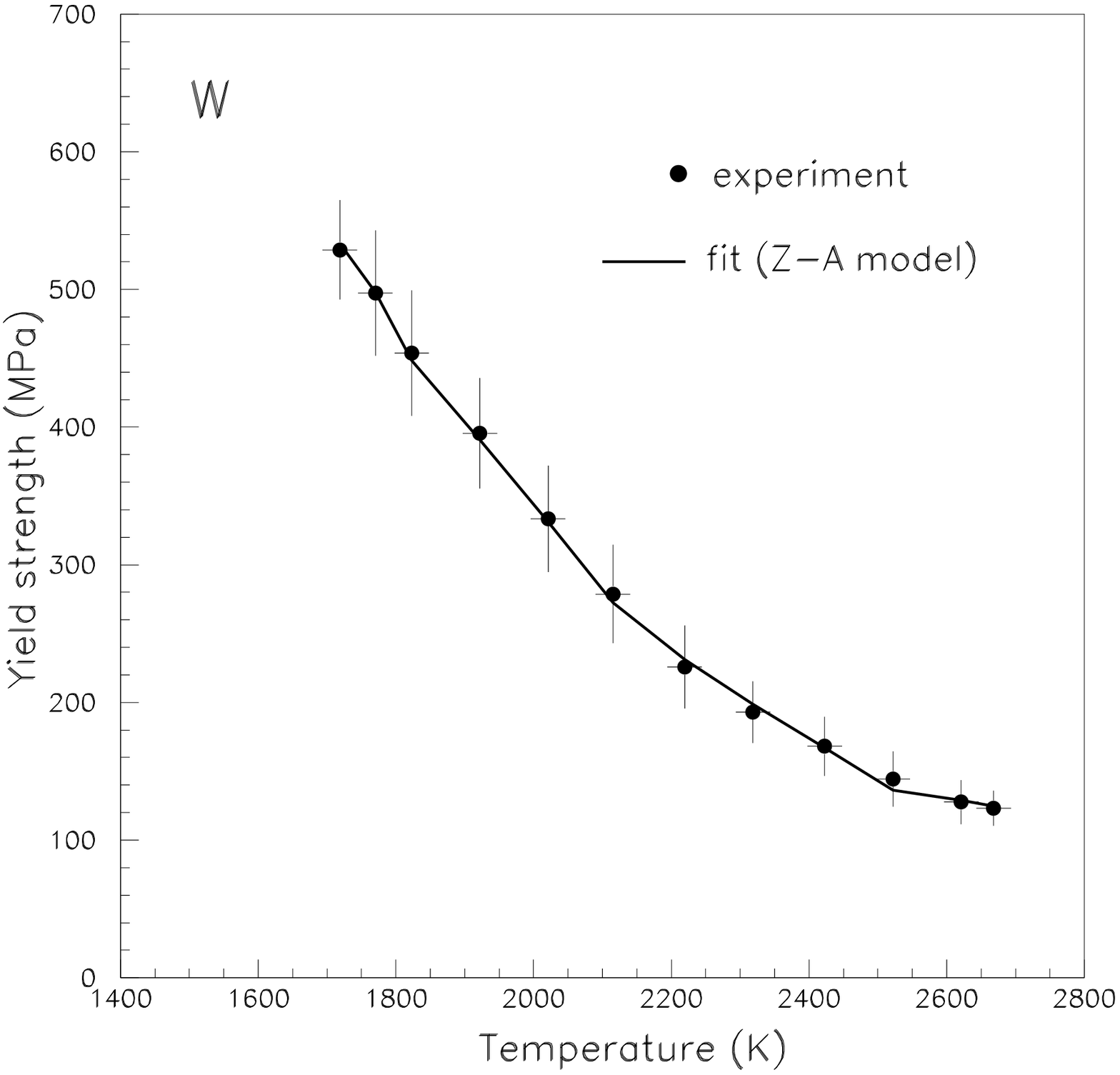}{}}

\vskip 0.3cm
{Figure 4. Experimental data on tungsten yield strength and the best fit
based on Z$\&$A model.}
\end{center}
\vskip 0.5cm

%------------------------------------------
\newpage
\vspace*{8.5cm}

\begin{center}
\hspace*{-9.cm}
\parbox{2cm}{\epsfxsize=2.cm \epsfysize=2.cm \epsfbox[00 00 90 90]
{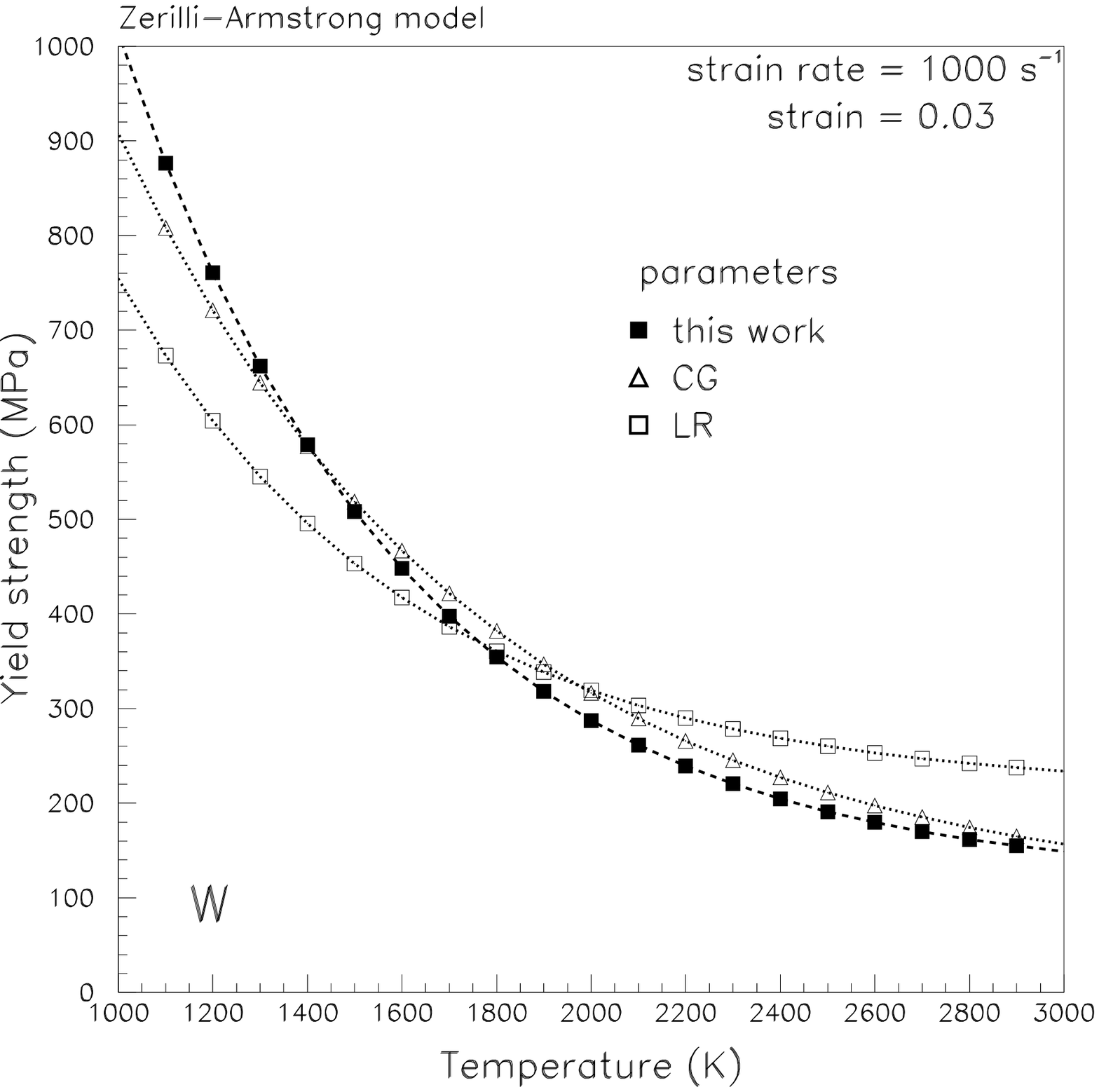}{}}

\vskip 0.3cm
{Figure 5. Tungsten yield strength, calculated using the Z$\&$A constitutive
equation, as a function of temperature
 for the strain rate of $1000\ s^{-1}$ and strain of $3\%$.
 Parameterizations 'CG' \cite{Ch95} and 'LR' \cite{Extr} are extrapolated to
 cover the temperature range explored in \cite{Nim}.}
\end{center}
\vskip 0.5cm

%------------------------------------------
\newpage
\vspace*{8.5cm}

\begin{center}
\hspace*{-9.cm}
\parbox{2cm}{\epsfxsize=2.cm \epsfysize=2.cm \epsfbox[00 00 90 90]
{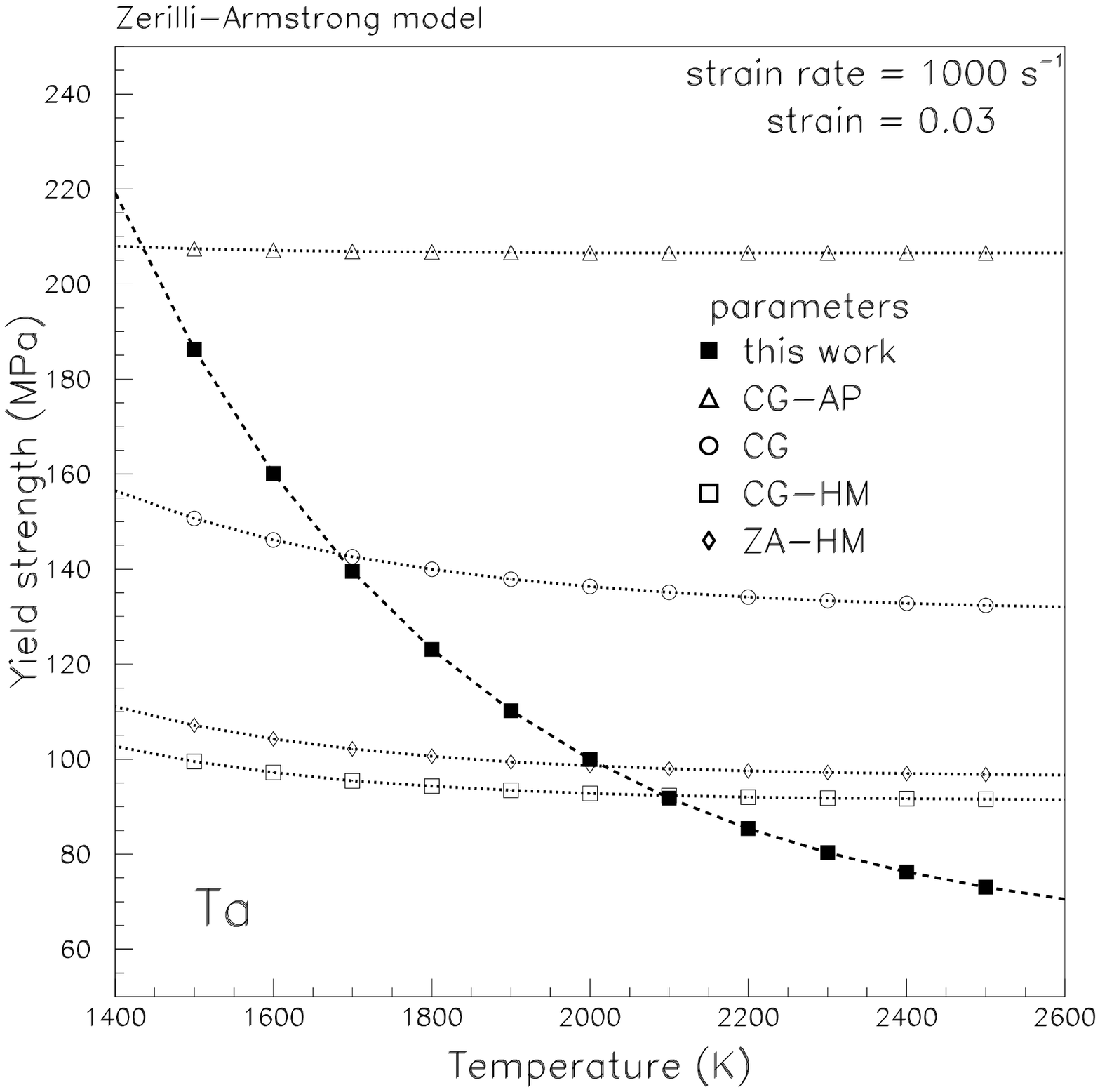}{}}

\vskip 0.3cm
{Figure 6. Tantalum yield strength, calculated using the Z$\&$A constitutive
equation, as a function of temperature
 for the strain rate of $1000\ s^{-1}$ and strain of $3\%$.
 Parameterizations 'CG-HM' \cite{Ch95}, 'CG' \cite{Ch95}, 'ZA-HM' \cite{ZA90} and 'CG-AP'\cite{Chgr}
 are extrapolated to cover the temperature range explored in \cite{Nim}.}
\end{center}
\vskip 0.5cm

\end{document}